\documentclass[a4paper,11pt]{article}
\usepackage[dvipdfmx]{graphicx}
\usepackage{xcolor}
\usepackage{jheppub} % for details on the use of the package, please
\usepackage{amsmath}    % required for `\split' (yatex added)
\usepackage[T1]{fontenc} % if needed
\renewcommand{\(}{\left(}
\renewcommand{\)}{\right)}
\newcommand{\bs}{\boldsymbol}
\newcommand{\df}{\dfrac}
\newcommand{\hc}{\text{h.c.}}
\newcommand{\sr}{\sqrt}
\newcommand{\cN}{\mathcal{N}}

\newcommand{\cL}{\mathcal{L}}
\newcommand{\pd}{\partial}

\newcommand{\nn}{\nonumber}
\newcommand{\ol}[1]{\overline{#1}}
\renewcommand{\a}{\alpha}
\renewcommand{\b}{\beta}

\renewcommand{\b}{\bar}
\newcommand{\der}{\partial}
\renewcommand{\d}{\dot}
\renewcommand{\ol}{\bar}

\newcommand{\calF}{{\cal F}}
\newcommand{\calG}{{\cal G}}
\newcommand{\Dg}{{\bs D}}
\newcommand{\wDg}{{\widehat {\bs D}}}

\title{Ghost-free vector superfield actions 
in supersymmetric higher-derivative theories}

\author[a]{Toshiaki Fujimori,}
\author[a]{Muneto Nitta,}
\author[a]{Keisuke Ohashi,}
\author[b]{Yusuke Yamada,}
\author[c]{Ryo Yokokura}

\affiliation[a]{Department of Physics 
\&
Research and Education Center for Natural Sciences,
Keio University,\\ {\small{\it Hiyoshi 4-1-1, Yokohama, Kanagawa 223-8521, Japan}}}
\affiliation[b]{SITP and Department of Physics, Stanford University, Stanford, California 94305, USA}
\affiliation[c]{Department of Physics, Keio University, 
Hiyoshi 3-14-1, Yokohama, Kanagawa 223-8522, Japan}

\emailAdd{toshiaki.fujimori018@gmail.com}
\emailAdd{nitta@phys-h.keio.ac.jp}
\emailAdd{keisuke084@gmail.com}
\emailAdd{yusukeyy@stanford.edu}
\emailAdd{ryokokur@rk.phys.keio.ac.jp}

\abstract{
We systematically construct ghost-free higher-derivative actions of Abelian vector supermultiplets in four-dimensional $\cN=1$ global supersymmetric theories. 
After giving a simple example which illustrates 
that a naive introduction of 
a higher-derivative term gives rise to a ghost, 
we discuss possible building blocks for a ghost-free action
and explicitly show that their bosonic parts have no ghost mode
and the auxiliary field $\bs{D}$ does not propagate.
Higher-derivative terms yield higher powers of the auxiliary field $\bs D$ in the actions, 
and the D-term equations of motion consequently
 admit multiple solutions in general. 
We confirm that the well-known 
supersymmetric Dirac-Born-Infeld action falls into this class.
We further give another example 
in which the standard quadratic kinetic term (Maxwell term) 
is corrected by a quartic term of the field strength.
We also discuss possible couplings to matter fields and a deformed D-term potential.
}

\begin{document} 
\maketitle
\flushbottom
\section{Introduction}
\label{intro}
Higher-derivative terms beyond renormalizable 
ones inevitably appear 
in effective theories of fundamental UV theories, 
such as unified theories with gravity, 
which would describe our universe. 
Once we consider such a nonrenormalizable theory, 
higher-derivative interactions appear which are absent in a renormalizable system. 
For example, the effective field theory of superstring theory contains such derivative interactions as well as non-derivative ones. 
However, we may need to be careful of higher-derivative interactions. 
As is well known, derivative interactions beyond second order can lead to a problem called the Ostrogradsky instability~\cite{Ostrogradski,Woodard:2006nt}, 
which causes a Hamiltonian unbounded from below. 
However, even when an interaction term contains more than two derivatives, 
some specific classes of such interactions do not exhibit the instability. 
One sufficient condition is that the equations of motion (E.O.M) in the system are at most second order differential equations. 
In a scalar-tensor system, 
the Horndeski class action is one example of a ghost-free higher-derivative system 
\cite{Horndeski:1974wa,Kobayashi:2011nu}. 
Recently, the higher-derivative action of a Proca field 
is also drawing attention in the context of cosmology 
\cite{Tasinato:2014eka,Heisenberg:2014rta,Allys:2015sht}. 
Although the instability itself is not a real problem 
if there is a ghost-free UV completion, 
the investigation of ghost-free higher-derivative interactions 
would be an interesting problem in its own right.

Supersymmetry (SUSY) was proposed to solve the hierarchy problem of the standard model and is one of important tools to investigate physics beyond the standard model. 
The effective theories of UV complete SUSY theories 
such as superstring theory 
naturally contain higher-derivative terms 
as corrections to the leading two-derivative terms.
However, higher-derivative terms in SUSY field theories often encounters the so-called auxiliary field problem 
\cite{Gates:1995fx-0,Gates:1995fx-1,Gates:1995fx-2,Gates:1995fx-3,Gates:1995fx-4}:
the action contains spacetime derivatives of the SUSY auxiliary fields 
($F$ and $\bs D$ for chiral and vector multiplets, respectively), 
so that one cannot eliminate them by their E.O.M. 
It was recognized~\cite{Gates:1995fx-0,Gates:1995fx-1,Gates:1995fx-2,Gates:1995fx-3,Gates:1995fx-4}
that this problem may occur 
in several higher-derivative chiral models 
such as a Wess-Zumino term~\cite{Nemeschansky:1984cd-0,Nemeschansky:1984cd-1},
and Skyrme~\cite{Bergshoeff:1984wb} and 
Faddeev-Skyrme~\cite{Freyhult:2003zb} models. 
In most cases, the derivatives on the auxiliary fields and the higher-derivative ghosts come up together~\cite{Antoniadis:2007xc}.\footnote{
If one can eliminate a ghost with a gauging by
introducing an auxiliary vector superfield, 
one can construct a model with a propagating ``auxiliary'' field without a ghost~\cite{Fujimori:2016udq}, which we may call a ghostbuster mechanism. }
Finally, a systematic classification of 
higher-derivative terms which are free from ghosts as well as the auxiliary field problem was given in 
Refs.\,\cite{Khoury:2010gb,Khoury:2011da,Koehn:2012ar,Koehn:2012te}.
Such terms were used in several SUSY 
higher-derivative chiral models; 
low-energy effective theory~\cite{Buchbinder:1994iw-0,Buchbinder:1994iw-1,Buchbinder:1994iw-2,Buchbinder:1994iw-3},  
coupling to supergravity~(SUGRA) 
\cite{Koehn:2012ar,Farakos:2012qu} 
and its applications~\cite{Khoury:2011da} 
to Galileons~\cite{Nicolis:2008in} 
and ghost condensation \cite{Koehn:2012te}, 
a Dirac-Born-Infeld (DBI) inflation~\cite{Sasaki:2012ka}, 
flattening of the inflaton potential~\cite{Aoki:2014pna-0,Aoki:2014pna-1},
topological solitons such as a BPS baby Skyrme model 
\cite{Adam:2013awa-0,Adam:2013awa-1,Nitta:2014pwa-0,Nitta:2014pwa-1,Bolognesi:2014ova-0,Bolognesi:2014ova-1}, 
a Skyrme model \cite{Gudnason:2015ryh-0,Gudnason:2015ryh-1,Queiruga:2015xka},
BPS solitons \cite{Nitta:2014pwa-0,Nitta:2014pwa-1,Queiruga:2017blc}
and their effective action \cite{Eto:2012qda}, 
nonlinear realizations \cite{Nitta:2014fca}, 
and a possibility of modulated vacua \cite{Nitta:2017yuf-0,Nitta:2017yuf-1}.
In addition, different ghost-free higher-derivative 
actions
of a chiral superfield 
are possible in the global~\cite{Farakos:2013zya} 
and SUGRA~\cite{Farakos:2012je} cases 
as well as
a non-local theory \cite{Kimura:2016irk}. 

On the other hand, there have been no such systematic studies for vector
multiplets so far.
There are only few examples of ghost-free actions 
of a vector multiplet.
One is the DBI action in 4D $\cN=1$ SUSY 
\cite{Cecotti:1986gb,Bagger:1996wp,Rocek:1997hi}. 
Since it is totally written in terms of the field strength 
$F_{mn}=\partial_m A_n-\partial_n A_m$ 
without any additional derivative operators, 
its E.O.M is of a second order, and no additional (ghost) mode appears.
The coupling to SUGRA has also been discussed in 
Refs.\,\cite{Cecotti:1986gb,Kuzenko:2002vk-0,Kuzenko:2002vk-1,Abe:2015nxa}. 
The SUSY Euler-Heisenberg action is one of ghost-free extensions of the Maxwell system~\cite{Farakos:2012qu,Cecotti:1986jy,Farakos:2013zsa,Dudas:2015vka}.
The others are 
a higher derivative extension of a scalar potential
in SUGRA~\cite{Cecotti:1986jy}, 
and a SUSY extension of non-linear self-dual 
actions for vector gauge fields \cite{Kuzenko:2002vk-0,Kuzenko:2002vk-1,Kuzenko:2000tg-0,Kuzenko:2000tg-1,Kuzenko:2000tg-2}. 

In this paper, we systematically study 
ghost-free higher-derivative actions of vector multiplets 
in 4D $\cN=1$ SUSY theories.
The known example of the DBI action gives us useful insights to find possible ghost-free interactions. 
To construct 
the most general ghost-free SUSY higher-derivative actions,
the nontrivial question is the conditions for the absence of the higher-derivative ghosts.
We will follow the following simple argument as a guiding principle: 
if the bosonic part of the action is ghost-free, 
the total system would be ghost-free as well.
This argument would be justified unless some fermions in the system condensate and acquire vacuum expectation values.\footnote{
This discussion is applied for bosonic ghosts and 
does not exclude a possibility of fermionic ghosts. 
For instance, 
a fermionic ghost exists when the vacuum energy becomes negative \cite{Nitta:2017yuf-0,Nitta:2017yuf-1}.
For a general discussion for fermionic ghosts, 
see Refs.~\cite{Villasenor:2002kw,Kimura:2017gcy}.
}
Another nontrivial point is the auxiliary field problem.
Typically, auxiliary fields become dynamical due to higher-derivative interactions, as mentioned above. 
In most known cases, 
the dynamical auxiliary field leads to a ghost (super)field. 
We will show that the ghost-free higher-derivative action 
does not have any propagating auxiliary field 
but has higher powers of the auxiliary field, yielding corrections to the D-term potential.
We will show an example of such a corrected D-term potential, which has nontrivial vacua due to the corrections.
We also give a matter coupling to higher-derivative vector multiplets. 

This paper is organized as follows. 
In Sec.\,\ref{ghostfull}, we show a nontrivial example which has a higher-derivative ghost for comparison. 
Then we move to the construction of ghost-free models in Sec.\,\ref{generalGF}. 
First, we discuss possible building blocks for ghost-free higher-derivative actions in Sec.\,\ref{BB}. 
With such ingredients, 
we find the general SUSY action whose bosonic part is free from ghosts. 
We also show the corresponding on-shell action.
In Sec.\,\ref{EX}, 
we show some concrete examples of ghost-free models, 
their bosonic actions, 
and the vacua realized by corrected D-term potentials.
We first show that the well-known SUSY DBI action is obtained as a particular example of our construction. 
We next propose a simple example.
In the example, 
there are two kinds of solutions of the E.O.M for the auxiliary D-term, 
which give so-called canonical and non-canonical branches. 
On the one hand, in the canonical branch,
the bosonic action consists of 
the standard quadratic kinetic term of the field strength 
(Maxwell term)
and a higher-derivative term of fourth order.
On the other hand, in the non-canonical branch,  
the action contains only a quartic higher-derivative term of the field strength
without the Maxwell term~\cite{Farakos:2012qu,Farakos:2013zsa}. 
In addition, we mention possible couplings to matter fields in Sec.\,\ref{MC}. 
We conclude in Sec.\,\ref{conclusion} with a brief discussion. 
In Appendix \ref{vs}, the convention of 
a vector superfield is summarized.
Throughout this paper, we will use the convention of 
Wess and Bagger~\cite{Wess:1992cp}.

\section{A higher-derivative action for a vector multiplet with a ghost}
\label{ghostfull}
Before going to ghost-free models, 
we show a nontrivial example of a higher-derivative action of 
a vector multiplet $V$ with a ghost mode for comparison.
The convention of the vector multiplet is summarized in Appendix~\ref{vs}.

We consider the following system,
\begin{align}
\cL = \left( \int d^2 \theta \, \frac{1}{4} W^\a W_\a + \text{h.c.} \right)
+ \int d^4 \theta \, f(D^\a W_\a).
\label{ghost}
\end{align}
Here, $W_\alpha = -\frac{1}{4}\b{D}^2 D_\alpha V$ 
is a gauge invariant chiral superfield, and
$f$ is a real function such that
its first derivative $f'$ is a non-constant function.\footnote{
If $f$ is a constant or a linear function, 
the last term in the Lagrangian vanishes.} 
To show the existence of a ghost, we rewrite the action as follows,
Since $\ol{D}^2(D^\a W_\a)=D^2(D^\a W_\a)=0$, 
$D^\a W_\a$ is a real linear superfield
and hence we can rewrite the action as
\begin{align}
\cL = \left( \int d^2 \theta \, \frac{1}{4} W^\a W_\a + \text{h.c.} \right)
+ \int d^4 \theta \, f(L) + \int d^4 \theta \, U( L - D^\a W_\a),
\end{align}
where $L$ and $U$ are a real linear and a real superfield, 
 respectively. 
Note that the real superfield $U$ is an additional vector 
superfield, and this action is invariant 
under the following gauge transformation for $U$,
\begin{align}
U \to U - \Sigma' - \ol \Sigma',
\label{gaugeU}
\end{align}
where $\Sigma'$ is a chiral superfield.
The variation of $U$ gives the constraint $L = D^\a W_\a$,
which reproduces the original action. 
If we instead perform a super-partial integral, we obtain
\begin{align}
\cL =& \int d^2 \theta \, \left( \frac{1}{4} W^\a W_\a + \frac{1}{2} \widehat{W}^\a W_\a + \text{h.c.} \right) + 
\int d^4 \theta \, \Big[ f(L) + U L \Big],
\label{eq:ghost_vector}
\end{align}
where we have used
\begin{align}
\int d^4 \theta \, U D^\a W_\a 
= - \int d^4 \theta \, D^\a U W_\a 
= - \int d^2 \theta \, \widehat{W}^\a W_a, 
\end{align}
with $\widehat{W}_\a=-\frac{1}{4}\ol{D}^2D_\a U$. 
By diagonalizing the kinetic terms of 
$W^\a$ and $\widehat{W}^\a$ in Eq.\,\eqref{eq:ghost_vector}, 
we find that one combination of the vector superfields 
has a kinetic term with a wrong sign
implying the existence of a ghost vector superfield.
Note that our action in Eq.~\eqref{ghost}
is a generalization of an action with a ghost given 
in Ref.~\cite{Dudas:2015vka}.

We may trade the linear superfield for
 a chiral superfield, which makes the situation clearer. 
The part of the action containing the linear superfield 
can be rewritten as 
\begin{align}
\int d^4 \theta \, f(\tilde L) + \int d^4\theta \, U \tilde{L} + 
\int d^4 \theta \, \tilde L ( \Phi + \ol \Phi ),
\end{align}
where $\tilde{L}$ is a general superfield and 
$\Phi$ is a chiral superfield.
The variation of $\Phi$ gives 
the constraint for a linear superfield 
$\ol{D}^2\tilde L=0$, which reproduces $\tilde{L}=L$,
whereas the variation of $\tilde L$ gives 
$f'(\tilde L) + (\Phi+\ol{\Phi}+U) = 0$.
As long as this equation can be solved (implicitly) as 
$\tilde{L} = \tilde{L}(\Phi+\ol\Phi+U)$, 
the action can be rewritten as
\begin{align}
\cL =& \int d^2 \theta \, \left( \frac{1}{4} W^\a W_\a + \frac{1}{2} \widehat{W}^\a W_\a + \text{h.c.} \right) + 
\int d^4 \theta \, g(\Phi+\ol\Phi+U),
\end{align}
where  
\begin{equation}
g(\Phi+\ol\Phi+U)=\left\{f(\tilde{L})+\tilde L(\Phi+\ol\Phi+U)\right\}\Biggr|_{\tilde L=\tilde L(\Phi+\ol\Phi+U)}. 
\end{equation}
Note that $\Phi$ is a St\"uckelberg
superfield 
which transforms as $\Phi\to \Phi+\Sigma'$ 
under the gauge transformation of $U$ in Eq.~\eqref{gaugeU}. 

Thus we have shown that the higher-derivative interaction in Eq.~\eqref{ghost} leads to a ghost vector superfield. 
Note that if $f'$ is a constant, 
the equation $f'(\tilde L)+(\Phi+\ol{\Phi}+U)=0$
can be solved with respect to $U$, which then
 becomes a composite superfield.
Consequently, no ghost fields arise.

\section{Ghost-free higher-derivative actions for vector multiplet}
\label{generalGF}
In this section, we discuss possible ghost-free 
higher-derivative actions of a vector multiplet. 
Since a massive vector superfield belongs to a reducible representation, 
we focus only on a massless gauge superfield. 
First, we consider building blocks for a higher-derivative action. 
To preserve the gauge symmetry, 
we need to use the gauge invariant superfield 
$W_\a=-\frac{1}{4}\ol{D}^2D_\a V$, 
which we call the ``gaugino superfield''. 
According to the Ostrogradsky's method, 
ghost modes appear if the equation of motion is
higher than second-order differential equations. 
Since the $\theta$-component of $W_\alpha$ 
contains the field strength $F_{mn}$, 
the spacetime derivatives of $W_\alpha$ in the superfield action would lead to higher-derivative terms in the E.O.M, 
and consequently give rise to ghost modes. 
For example, a simple interaction with spacetime derivatives
$\int d^2\theta \, \pd^m W^\a\pd_m W_\a$ 
leads to a fourth-order differential equation 
for the gauge field $A_m$. 

\subsection{The building blocks of ghost-free higher-derivative terms}
\label{BB}
In this subsection, 
we write down possible building blocks of higher-derivative actions.
As we mentioned in Sec.\,\ref{intro}, 
we only focus on purely bosonic parts of higher-derivative actions 
and require them to be ghost-free.
We propose the following three simple conditions 
to realize ghost-free bosonic terms:
\begin{itemize}
\item 
To obtain a manifestly gauge invariant action, 
we require that the integrand of the superfield action is 
written in terms of the gaugino superfield $W_\alpha$ 
rather than the vector superfield $V$. 
Consequently, the bosonic part of the action is  
written totally in terms of the field strength, which is manifestly gauge invariant.

\item 
The Lagrangian has
no spacetime derivatives of the field strengths
such as $\der_m F_{np}$, 
which may lead to ghosts.\footnote{
Here, we do not mean the absence of the terms 
e.g. $A^m \partial^n F_{mn}$. 
This conflicts with the first requirement 
that the Lagrangian is a function of field strength.}

\item 
SUSY invariants which we focus on should have 
at least one purely bosonic term. 
Only such terms are relevant for the presence/ absence of ghosts. 
\end{itemize}

Taking into account these criteria, 
we find the following most general form of 
the higher-derivative Lagrangians:
\begin{equation}
 {\cal L}
=
\left( \int d^2\theta \, \calF(H) \, W^2 + \text{h.c.} \right)
+
\int d^4\theta \, \calG (H,\b H, \wDg) \, W^2\b{W}^2,
\label{hd}
\end{equation}
where (anti-)chiral superfield  $H (\b H)$ and  real superfield $\wDg$ are defined by 
\begin{eqnarray}
H \equiv -\frac{1}{4} \bar D^2 \bar W^2, \hspace{7mm} 
\b H\equiv -\frac{1}{4}  D^2 W^2, \hspace{7mm} 
\wDg \equiv -\frac{1}{2} D^\alpha W_\alpha 
= -\frac{1}{2} \bar D_{\dot \alpha} \bar W^{\dot \alpha}.
\end{eqnarray} 
Here $\calF$ and $\calG$ are arbitrary functions of arguments, 
which are holomorphic and real scalar superfields, respectively.
In the following, we explain how the Lagrangian in Eq.\,\eqref{hd}
can be obtained from the above three conditions.

Firstly, we focus on the ``irreducible'' components 
of the gaugino superfield, which is not made of other (lower) components of a supermultiplet.
In the gaugino superfield, 
there is only one purely bosonic irreducible component\footnote{
Note that $F_{mn}$ is not an irreducible component in the vector
superfield $V$. However, we can regard $F_{mn}$ as an irreducible 
component in the gaugino superfield $W_\alpha$.} 
\begin{equation}
D_\alpha W_\beta|=
-i(\sigma^{mn}\epsilon)_{\alpha\beta} F_{mn}
-\epsilon_{\alpha\beta}\bs{D},
\end{equation}
where the vertical bar denotes 
$\theta =\b\theta =0$ projection on a superfield,
and 
$(\sigma^{mn}\epsilon)_{\alpha\beta}
=\frac{1}{4}
(\sigma^m_{\alpha\d\gamma}\b\sigma^{n\d\gamma \delta}
-
\sigma^n_{\alpha\d\gamma}\b\sigma^{m\d\gamma \delta})
\epsilon_{\delta\beta}$.
This implies that 
 the numbers of derivatives 
$D_\alpha$ and $\b{D}^{\d\alpha}$ 
in the component action
should be equal to those of the gaugino superfield 
$W_\alpha$ and $\b{W}^{\d\alpha}$, respectively. 
In other words, the bosonic part ${\cal L}_b $ of 
the Lagrangian should take the form of 
\begin{equation}
{\cal L}_b
= {\cal L}_b (D_\alpha W_\beta|, \b{D}^{\d\alpha}\b{W}^{\d\beta}|).
\label{bos}
\end{equation}
Furthermore, the Lorentz invariance requires 
the Lagrangian ${\cal L}_b$ to be 
composed of Lorentz invariant combinations of 
$D_\alpha W_\beta|$ and $\b{D}^{\d\alpha}\b{W}^{\d\beta}|$.
We can show that the following three Lorentz invariant combinations 
are independent building blocks of the Lagrangian
\begin{equation}
D^\alpha W_\alpha, \hspace{5mm} 
D^\alpha W^\beta D_\alpha W_\beta, \hspace{5mm}  
\b{D}_{\d\alpha}\b{W}_{\d\beta} \b{D}^{\d\alpha}\b{W}^{\d\beta}.
\label{3bb}
\end{equation}
This follows from the irreducible decomposition of $D_\alpha W_\beta$ 
\begin{equation}
 D_\alpha W_\beta = \df{1}{2} 
\epsilon_{\alpha\beta} D^\gamma W_\gamma 
+
D_{(\alpha} W_{\beta)},
\end{equation}
and the relation
\begin{equation}
 D^{(\alpha} W^{\beta)}
D_{(\beta} W_{\gamma)}
=
\delta^\alpha_\gamma
\(
\df{1}{2} D^\beta W^\delta D_\beta W_\delta 
-\df{1}{4}(D^\beta W_\beta)^2
\).
\end{equation}
Here, the parenthesis for spinors means the symmetrization:
$D_{(\alpha}W_{\beta)}=
\frac{1}{2}(D_\alpha W_\beta +D_\beta W_\alpha)$.
Similar equations hold for $\b{D}^{\d\alpha} \b{W}^{\d\beta}$.
Thus we can express Lorentz invariants given by 
contractions of the spinors of $D_\alpha W_\beta$ and 
$\b{D}^{\d\alpha }\b{W}^{\d\beta}$ in terms of \eqref{3bb}.
\footnote{Note that there are ambiguities of choosing 
the independent quantities. 
For example, we can choose $D^\alpha W^\beta D_\beta W_\alpha$ 
instead of $D^\alpha W^\beta D_\alpha W_\beta$.
However, the number of the independent quantities does not change.} 
Note that $D^\alpha W_\alpha=-2\wDg$ is a real superfield. 
We can replace $D^\alpha W^\beta D_\alpha W_\beta$ 
with $\b H$ since   
\begin{equation}
D^\alpha W^\beta{D}_{\alpha}{W}_{\beta}
=2\b H+({D}^2 {W}_{\alpha}) {W}^{\alpha},
\end{equation}
and the last term has only fermionic terms, 
which are irrelevant to our discussion. 
Therefore, the Lagrangian should take the form
\begin{eqnarray}
{\cal L}_b={\cal L}_b(\bar J\equiv H\big|,J\equiv \b H \big|, \Dg=\wDg \big|), \label{BL}
\end{eqnarray} 
where 
\begin{equation}
J:= -\df{1}{4}D^2W^2|
=
-\dfrac{1}{2} F^{mn} F_{mn} +
\dfrac{i}{2} F_{mn} \tilde F^{mn} 
+ \boldsymbol{D}^2
- 2 i \lambda^\alpha (\bar\sigma^m)_{\alpha\d\beta} 
\partial_m \bar\lambda^{\d\beta},
\end{equation}
and $\tilde{F}^{mn} = \frac{1}{2} \epsilon^{mnpq}F_{pq}$. 
The form \eqref{BL} simply implies 
that the Lagrangian should be 
a function of $(F_{mn}F^{mn}, F_{mn}\tilde{F}^{mn}, \bs D)$. 

In order to embed the bosonic expression \eqref{BL} to a SUSY system, 
one needs to describe it 
as a superspace integration of superfields.
Since $\int d^2\theta $ and $\int d^4\theta $ integrations are equivalent to acting two and four spinor derivatives on superfields, 
the super-integrands have to be 
respectively proportional to 
$W_\alpha W_\beta$ for $\int d^2\theta $ integrals and 
$W_\alpha W_\beta \b{W}_{\d\alpha}\b{W}_{\d\beta}$ 
for $\int d^4 \theta$ integrals 
so that the resulting component action 
takes the form ${\cal L} (H,\bar H, \bs D)$. 
From these observations, 
we obtain the most general Lagrangian 
satisfying all the conditions:\footnote{
Note that $W_\alpha W_\beta$ can be rewritten as $W_\alpha W_\beta=\frac{1}{2}\epsilon_{\alpha\beta} W^2$.
}
\begin{equation}
{\cal L}
=
\left(
\int d^2\theta \, \calF(H) \, W^2
+\text{h.c.}
\right)
+
\int d^4\theta \, \calG(H,\b H,\wDg) \, W^2\b{W}^2,
\end{equation}
where $\calF$ and $\calG$ are 
arbitrary holomorphic and real scalar functions, respectively. 

We note that $\wDg\propto D^\alpha W_\alpha$ dependence 
causes no ghost mode in contrast to the previous section 
since $\calG \, W^2 \bar W^2$ is not invertible 
with respect to $D^\alpha W_\alpha$. 
Then, repeating the procedure discussed 
in the previous section, 
we find out that the vector superfield $U$ satisfying 
$\calG' W^2\bar W^2+\Phi+\bar \Phi+U=0$ 
is always a composite superfield.
Therefore, the function $\calG$ is allowed to be a function of $\wDg$. 

Among $H$, $\b H$, and $\wDg$, only $H$ is a chiral superfield,
and hence the chiral superfield $\calF$ needs to be 
a holomorphic function of $H$.
As long as $\calF$ is a smooth function around $H=0$, 
it can always be written as
\begin{align}
\calF(H) = \tau + h(H)H,
\end{align}
where $\tau$ is a constant and $h(H)$ is a holomorphic function.  
The second term can be absorbed into $\calG$ due to the identity
\begin{eqnarray}
\int d^2\theta \, h(H) H W^2 = 
-\frac{1}{4} \int d^2 \theta \, \b D^2 \left( h(H) \b W^2 W^2 \right)
= \int d^4\theta \, h(H) W^2 \b W^2.
\end{eqnarray}
Therefore we can set $\calF = \tau = \text{const}$. 
Consequently, the bosonic part of Lagrangian takes the form
\begin{eqnarray}
{\cal L}_b = \left( \tau J + {\rm h.c.} \right) + \calG (\b J, J,\Dg) |J|^2. \label{bosonicLag}
\end{eqnarray}
The first term gives the standard quadratic terms 
(Maxwell, theta and D-terms), 
while the second term is the higher-derivative correction, 
which is required to be proportional to $|J|^2$ 
due to SUSY.\footnote{
Thanks to this property, 
the self-dual (instanton) solution of the Yang-Mills equation 
would not be modified by the additional higher-derivative terms
in the non-Abelian extension of our model.} 

Finally, we note that there is no derivative acting on the auxiliary field $\bs D$ in our general action, and hence the solution of its E.O.M can be expressed as a local function of physical fields, 
i.e. $F_{mn}$ without derivatives on it.\footnote{
In the presence of matter fields, 
$\bs D$ can also depend on matter fields without derivatives.} 
Therefore, if we assign an appropriate sign for the quadratic term, 
the on-shell action is always ghost-free as we expected.

\subsection{Examples of ghost-free higher-derivative action 
for vector multiplet}
\label{EX}
In the previous subsection, we have derived the general SUSY action, which has a ghost-free purely bosonic part.
In this section, we consider some specific examples. 
We can construct various models by choosing $\calF$ and $\calG$ 
in our general action \eqref{hd}. 

First, let us choose $\calF=1/4g^2$ and
$\calG=\calG(D^\alpha W_\alpha)$:
\begin{equation}
 {\cal L}
=
\(\df{1}{4g^2}\int d^2\theta W^2 + \hc\) +
\int d^4\theta \, \calG(D^\alpha W_\alpha) W^2\b{W}^2.
\end{equation}
Here, $g$ is a coupling constant.
Note that the reality of $\calG$ is satisfied 
because $D^\alpha W_\alpha$ is a real superfield 
$D^\alpha W_\alpha =\b{D}_{\d\alpha}\b{W}^{\d\alpha}$. 
Since $D^\alpha W_\alpha|=-2\bs{D}$, to the bosonic part, 
$\calG$ only gives higher-order terms of auxiliary fields, i.e. 
\begin{equation}
{\cal L}_b = 
\df{1}{4g^2}(J+\b{J})+
\calG(-2\bs{D}) |J|^2. 
\label{eq:former}
\end{equation}

Second, for $\calG=\calG(H,\b H)$, 
the corresponding action is given by
\begin{equation}
{\cal L} = 
\(\df{1}{4g^2}\int d^2\theta W^2 + \hc\) + 
\int d^4 \theta \, \calG(H,\b{H})W^2\b{W}^2.
\label{D2W2}
\end{equation}
In this case, the bosonic action is 
\begin{equation}
{\cal L}_b = 
\df{1}{4g^2}(J+\b{J})+\calG(\b{J},J)|J|^2.
\label{gba}
\end{equation}
Contrary to the previous example, 
this action contains 
higher order terms of the field strength $F_{mn}$. 
This difference would be useful for model buildings. 

Obviously, these action leads to a second order E.O.M for the vector field. 
Furthermore, there is no derivative interaction of the auxiliary field $\boldsymbol{D}$ as we mentioned. 
Hence, the system does not have any additional degrees of freedom. 
Since the number of bosonic degrees of freedom does not change, 
that of fermions also remains the same due to SUSY.

The DBI action is a well-known example of 
a ghost-free higher-derivative action.
It is realized with the following set of choices
\cite{Cecotti:1986gb,Bagger:1996wp},
\begin{align}
&\calF = \frac{1}{4g^2}= \frac{\mu \alpha^2}4,\quad {\rm with~} \alpha \equiv \frac1{\sqrt{\mu g^2}} \nn\\
&\calG=\frac{\mu \alpha^4 }4\left(1-\frac{\alpha^2}{4}({H}+\ol{H})+\sqrt{1-\frac{\alpha^2}2 ({H}+\ol{H})+\frac{\alpha^4}{16}({H}-\ol{{H}})^2}\right)^{-1}.
\end{align}
The corresponding bosonic Lagrangian is given by 
\begin{eqnarray}
\cL_b&=& 
\df{\mu \alpha^2}{4} (J+\b J) +\frac{\mu \alpha^4}4 \left(1-\frac{\alpha^2}{4}({J}+\ol{J})+\sqrt{1-\frac{\alpha^2}2 ({J}+\ol{J})+\frac{\alpha^4}{16}({J}-\ol{{J}})^2}\right)^{-1}|J|^2
\nn \\
&=&\mu -\mu \sqrt{1-\frac{\alpha^2}2 ({J}+\ol{J})+\frac{\alpha^4}{16}({J}-\ol{{J}})^2}\nn \\
&=&\mu -\mu \sqrt{1+\frac{\alpha^2}2 F_{mn}F^{mn}-\alpha^2{\bs D}^2-\frac{\alpha^4}{16}(\tilde F^{mn}F_{mn})^2}\nn \\
&=&\mu -\mu \sqrt{-\text{det}(\eta_{mn}+\alpha F_{mn})-\alpha^2 {\bs D}^2}. \phantom{\Bigg[}
\end{eqnarray}
We find that $\bs{D}=0$ is the unique solution of E.O.M for $\bs{D}$.
Thus we obtain the on-shell action
\begin{align}
\int d^4x\, \mu 
\biggl(- \sqrt{-\text{det}(\eta_{mn}+\alpha F_{mn})}+1\biggr).
\end{align}

Next, let us consider the following simplest 
example~\cite{Cecotti:1986jy,Farakos:2012qu,Farakos:2013zsa,Dudas:2015vka}:
\begin{equation}
S=\df{1}{4} \int d^4x d^2\theta 
\( \frac{1}{g^2} + \df{1}{4\Lambda} \b{D}^2 \b{W}^2 \) W^2 + \hc
\label{hds}
\end{equation}
Here, $\Lambda$ is a real 
parameter of mass dimension four.
This action is realized by choosing $\mathcal F$ as
\begin{equation}
\mathcal F = \frac{1}{4g^2} + \df{1}{16\Lambda} \b{D}^2 \b{W}^2.
\end{equation}
The bosonic part of the action $S_b$ can be calculated as
\begin{equation}
S_b= 
\int d^4x 
\left[
-\df{1}{4g^2}  F^{mn}F_{mn} +\df{1}{2g^2}\bs{D}^2
-\df{1}{2\Lambda}\(-\df{1}{2}F^{mn}F_{mn}+\bs{D}^2\)^2
-\df{1}{2\Lambda}\(\df{1}{2}F^{mn}\tilde{F}_{mn}\)^2
\right].
\end{equation}
The E.O.M for the auxiliary field $\bs{D}$ reads:
\begin{equation}
0=\bs{D} 
\left[ 
\frac{1}{g^2}+\frac{2}{\Lambda}\(\df{1}{2}F^{mn}F_{mn}-\bs{D}^2\)
\right].
\end{equation}
Thus, we have the following three solutions for the auxiliary field: 
\begin{equation}
 \bs{D}=0, \quad
\pm \sr{\df{1}{2}\(F^{mn}F_{mn}+\df{\Lambda}{g^2}\)}.
\end{equation}
The on-shell action in the corresponding branches 
can be written as follows. 
\begin{itemize}
 \item The canonical branch (the case of $\bs{D}=0$):
\begin{equation}
S_b= 
\int d^4x 
\left[
-\df{1}{4g^2} F^{mn}F_{mn} 
-\df{1}{8\Lambda}\(F^{mn}F_{mn}\)^2
-\df{1}{8\Lambda}\(F^{mn}\tilde{F}_{mn}\)^2
\right].
\end{equation}
This action contains the higher-derivative terms 
$(F^{mn}F_{mn})^2$ and $(F^{mn}\tilde{F}_{mn})^2$ 
as well as the standard quadratic term $F^{mn}F_{mn}$. 
In this case, 
the higher-order terms vanish in the limit $|\Lambda|\to \infty$.

\item The non-canonical branch 
(the case of $\bs{D} = \pm \sqrt{(F^{mn}F_{mn} + \Lambda/g^2)/2}\ $):
\begin{equation}
S=\int d^4x \left[ 
-\df{1}{8\Lambda}(F^{mn}\tilde{F}_{mn})^2 +\df{\Lambda}{8g^4} 
\right].
\label{eq:non-canonical}
\end{equation}
An interesting property of this action is 
the absence of the quadratic kinetic term. 
Therefore, we call this branch the non-canonical one.
Since $\bs{D}$ is real, $F^{mn}F_{mn} +\Lambda/g^2$ 
should be non-negative.
Therefore the solution of non-canonical branch is 
consistent only when 
\begin{equation}
 F^{mn}F_{mn} + \df{\Lambda}{g^2} \geq 0.
\label{nccond}
\end{equation}
This inequality shows that 
$F_{mn}=0$ can be a consistent solution of the E.O.M
only when the parameter $\Lambda$ is positive.\footnote
{The case in which $\Lambda$ is negative is investigated in 
Ref.~\cite{Farakos:2013zsa}.
In such a case, $F^{mn}F_{mn}$ 
cannot become zero from the condition in Eq.~\eqref{nccond}. }
Note that this branch becomes 
ill-defined in the limit $|\Lambda|\to\infty$.
\end{itemize}

As shown above, 
there exist multiple branches
due to the higher-order terms of the auxiliary field $\bs D$. 
The presence of the multiple branches has been known 
in the case of chiral multiplets $\Phi$ with the action
$\sim \int d^4\theta \, |D^\alpha \Phi D_\alpha\Phi|^2$, 
where the SUSY higher-derivative term involves 
a quartic interaction of the F-term \cite{Adam:2013awa-0,Adam:2013awa-1,Nitta:2014pwa-0,Nitta:2014pwa-1}. 
Similarly to the present case, 
the F-term has three solutions, 
and they lead to a canonical branch with the standard kinetic term
and non-canonical branches without a second-order derivative term.

\section{A coupling to matter fields}
\label{MC}
So far, we have discussed higher-derivative actions 
composed purely of vector superfields.
To discuss more general and phenomenological models, 
we consider the following higher-derivative system 
coupled to the matter chiral superfields $\Phi_1$ and $\Phi_2$: 
\begin{align}
 {\cal L}=& \df{1}{2} \int d^4\theta 
\biggl( \b{\Phi}_1 \, e^{V} \Phi_1 + \b{\Phi}_2 \, e^{-V} \Phi_2 
+ {\cal G}(H,\b H,\wDg,\Phi_1,\bar \Phi_1e^{V},\Phi_2,\bar \Phi_2e^{-V})
\, W^2\b{W}^2+\xi V\biggr)
\nn\\
& + \df{1}{4g^2} \int d^2\theta \, W^2 + \hc,
\end{align}
where
$\xi$ is a real parameter called 
the Feyet-Iliopoulos (FI) parameter.
The bosonic part of the Lagrangian is given by 
\begin{equation}
\begin{split}
{\cal L}_b
&=
-{\cal D}^m A_1 {\cal D}_m A^*_1
-{\cal D}^m A_2 {\cal D}_m A^*_2
+
A_1 A^*_1 \bs{ D}
-
A_2 A^*_2 \bs{ D} 
+F_1F^*_1+F_2F^*_2
+\xi \bs{D}
\\
&\quad
-\df{1}{4g^2}F^{mn}F_{mn}
+\df{1}{2g^2}\bs{D}^2
+
{\cal G}( \b J, J, \Dg, 
A_{i}, A_{i}^*)|J|^2, 
\end{split}
\end{equation}
where $A_i=\Phi_i|$ ($i=1,2$) are complex scalar fields, 
and $F_i=-\frac{1}{4} D^2 \Phi_i$ are auxiliary fields 
for the chiral superfields, and 
${\cal D}_m$ is a gauge covariant vector derivative. 
Here, 
We find that there are no derivatives acting on the auxiliary field $\bs{D}$ as well as the auxiliary fields $F_i$ for the chiral superfields $\Phi_i$. 
Therefore, the solutions to the E.O.M for $\bs{D}$ and $F_i$ 
are functions of $A_i$, $A_i^*$ and $F_{mn}$
without spacetime derivatives acting on them. 
Thus, there are no ghost modes even when the vector superfield is coupled to matter fields.
 
\section{Summary and discussion}
\label{conclusion}
We have investigated ghost-free higher-derivative actions 
of a vector multiplet 
in 4D $\cN=1$ SUSY models. 
As we have shown in Sec.\,\ref{ghostfull}, a naive higher-derivative extension leads to a ghost mode, 
even though it preserves the gauge symmetry. 
We have taken rather general conditions 
that lead to the ghost-free bosonic action: 
the bosonic part of the Lagrangian must be 
a function of $F_{mn}$ and $\bs D$ 
without derivatives acting on them. 
In addition, it should be expressed by 
SUSY invariants. 
These simple requirements lead to 
the general action given in Eq.\,\eqref{hd}, 
which is characterized by a holomorphic function 
${\cal F}={\cal F}(\b{D}^2\b{W}^2)$ 
and a real function ${\cal G}$ 
of $D^\alpha W_\alpha$, $\b{D}^2\b{W}^2$ and $D^2 W^2$.
We have also shown some examples with specific ${\cal F}$ and ${\cal G}$. 
As we mentioned, 
the SUSY DBI action is one of particular known examples in our framework. 
In general, higher-derivative terms yield higher powers of the auxiliary field $\bs D$, and consequently the E.O.M for $\bs D$ admits multiple solutions.
In the other example in Eq.~\eqref{hds}, 
the E.O.M for $\bs D$ gives two types of solutions: 
the canonical branch, 
in which the bosonic action consists of 
the standard quadratic kinetic term of the field strength 
(the Maxwell term)
and a fourth order higher-derivative term,
and the non-canonical branch,  
which is composed only of a fourth order higher-derivative term 
without the standard kinetic term. 
We have also shown that the matter coupled case discussed in Sec.\,\ref{MC} still satisfies our criteria for the absence of ghosts. 
The resultant action would have a deformed D-term potential, which has potentially interesting applications to model buildings.

There are many possible applications of our results. 
Since it is straightforward to extend the action \eqref{hd} to SUGRA, 
we can consider the inflationary model buildings. 
Indeed, the DBI extension of a matter-coupled vector superfield action leads to a scalar potential flattened by higher-order terms~\cite{Abe:2015fha}. 
It would be interesting to consider the SUSY breaking due to 
the higher-dimensional operators~\cite{Farakos:2013zsa,Kuzenko:2011ti-0,Kuzenko:2011ti-1}. 
It may also be possible to apply our framework 
to construct effective 4D models of superstring, 
where the higher-order D-term plays an important role, 
since the higher-order F-term contribution has been applied 
to effective string models of moduli stabilization \cite{Ciupke:2015msa}
as well as inflation~\cite{Broy:2015zba}.

Since a non-canonical branch of 
the E.O.M for the auxiliary field
in higher-derivative chiral models gives rise to 
BPS equations for BPS baby Skyrmions 
\cite{Adam:2013awa-0,Adam:2013awa-1,Nitta:2014pwa-0,Nitta:2014pwa-1},
the non-canonical branch of 
our vector superfield in Eq.~\eqref{eq:non-canonical}
may give a new kind of BPS topological solitons or instantons.
It is also interesting to look for new BPS solitons 
in the matter coupled system discussed in Sec.\,\ref{MC}.

The generalization to a non-Abelian case is straightforward 
but interesting for various applications, e.g., in inflation 
\cite{Maleknejad:2011jw-0,Maleknejad:2011jw-1,Maleknejad:2011jw-2}.
In such a case, the higher-order action should be further restricted by the non-Abelian symmetry. 
In addition, our systematic analysis would be applicable to the construction of a ghost-free action of a linear superfield, 
which was studied in a few cases~\cite{Bagger:1997pi,Rocek:1997hi,Aoki:2016cnw}. 
As a generalization, 
we can also consider the combination of the ghost-free higher-derivative actions of matter fields and higher curvature in SUGRA. 
It would also be interesting to see whether compositions of ghost-free models remain ghost-free or not. 
We will address these issues elsewhere. 
Finally, in certain chiral models,
a ghost chiral superfield can be gauged out 
by introducing an auxiliary vector superfield without 
a kinetic term \cite{Fujimori:2016udq}.
It is an open question whether the ghost in a vector superfield 
discussed in Sec.\,\ref{ghostfull} can be eliminated 
by introducing an auxiliary two-form tensor gauge field allowing a  gauge transformation of a two-form tensor.

\section*{Acknowledgement}
This work is supported by  the Ministry of Education,
Culture, Sports, Science (MEXT)-Supported Program for the Strategic
Research Foundation at Private Universities ``Topological Science''
(Grant No.~S1511006).
The work of M.~N.~is also supported in part by a Grant-in-Aid for
Scientific Research on Innovative Areas ``Topological Materials
Science'' (KAKENHI Grant No.~15H05855) from the MEXT of Japan, and 
by the Japan Society for the Promotion of Science
(JSPS) Grant-in-Aid for Scientific Research (KAKENHI Grant
No.~16H03984).
The work of R.~Y.~is 
supported by Research Fellowships of JSPS
 for Young Scientists Grant Number 16J03226.
Y.~Y.~is supported by SITP and by the NSF Grant PHY-1720397.

\appendix
\section{Convention for vector superfield}
\label{vs}
In this appendix, we summarize our convention 
for vector superfields.
The convention is the same as Ref.\,\cite{Wess:1992cp}.

We denote a real vector superfield as $V$.
The gauge field $A_m$ 
is embedded into the vector component of $V$ as
\begin{equation}
 A_m=-\df{1}{4}\b\sigma^{\d\alpha \alpha}_m [D_\alpha,\b{D}_{\d\alpha}]V.
\end{equation}
The vector superfield $V$ transforms 
under a $U(1)$ gauge transformation as
\begin{equation}
V \to V+\Sigma +\b\Sigma,
\end{equation}
where $\Sigma$ is a chiral superfield $\b{D}_{\d\alpha} \Sigma=0$.
The gaugino superfield is defined by 
\begin{equation}
 W_\alpha =-\df{1}{4}\b{D}^2 D_\alpha V,
\end{equation}
which is invariant under the $U(1)$ gauge transformation.
The component fields in the gaugino superfield can be expressed as
\begin{equation}
 W_\alpha|= -\df{1}{4} \b{D}^2 D_\alpha V|= -i\lambda_\alpha,
\end{equation}
\begin{equation}
 D_\alpha W_\beta |
=-\df{1}{4}D_\alpha \b{D}^2 D_\beta V|
=
-i(\sigma^{mn}\epsilon)_{\alpha\beta} F_{mn}
 -\epsilon_{\alpha\beta}\bs{D}, 
\end{equation}
where $(\sigma^{mn}\epsilon)_{\alpha\beta}
=\frac{1}{4}
(\sigma^m_{\alpha\d\gamma}\b\sigma^{n\d\gamma \delta}
-
\sigma^n_{\alpha\d\gamma}\b\sigma^{m\d\gamma \delta})
\epsilon_{\delta\beta}$
and $F_{mn}=\der_m A_n -\der_n A_m$
is the field strength of the gauge field. 
Alternatively, $F_{mn}$ and $\bs{D}$ can be expressed as 
\begin{equation}
 F_{mn}=\df{1}{2i}
\((\sigma_{mn})_\alpha {}^\beta D^\alpha W_\beta
-(\b\sigma_{mn})^{\d\alpha} {}_{\d\beta}\b{D}_{\d\alpha}
 \b{W}^{\d\beta}\)|,
\end{equation}
\begin{equation}
 \bs{D}= -\df{1}{2}D^\alpha W_\alpha| =-\df{1}{2}\b{D}_{\d\alpha}\b{W}^{\d\alpha}|.
\end{equation}
Note that there are other component fields 
in $V$: $V|=C$, $D_\alpha V| =\chi_\alpha$,
$\b{D}^{\d\alpha} V| =\b\chi^{\d\alpha}$, 
$-\frac{1}{4}D^2 V|= M$, $-\frac{1}{4}\b{D}^2 V|=N$.
These components can be fixed to zero 
by imposing the Wess--Zumino gauge fixing condition,
and do not appear in component actions.
The ordinary kinetic term 
for the gauge superfield is given by (up to total derivative)
\begin{equation}
 {\cal L}
=\df{1}{4g^2}\int d^2\theta W^2 +\hc
=\df{1}{4g^2} J+\hc
=
-\dfrac{1}{4g^2} F^{mn}F_{mn} 
+\df{1}{2g^2} \boldsymbol{D}^2
-\df{i}{g^2}\lambda^\alpha (\bar\sigma^m)_{\alpha\d\beta} 
\partial_m \bar\lambda^{\d\beta},
\end{equation}
where $\int d^2\theta =-\frac{1}{4}D^2$, and
\begin{equation}
  J:= -\df{1}{4}D^2W^2|
=
-\dfrac{1}{2}F^{mn}F_{mn} +\dfrac{i}{2}
F_{mn}\tilde F^{mn}+\boldsymbol{D}^2
-2i\lambda^\alpha (\bar\sigma^m)_{\alpha\d\beta}
 \partial_m \bar\lambda^{\d\beta}.
\end{equation}

%%%%%%%%%%%%%%%%%%%%%%%%%%%%%%%%%%%%%%%%%%%%%


\begin{thebibliography}{99}
\bibitem{Ostrogradski}
  M.~Ostrogradsky, ``Memoires sur les equations differentielles relatives au probleme des
  isoperimetres,'' 
  Mem. \ Ac. \ St. Petersbourg VI, 385 (1850).

\bibitem{Woodard:2006nt} 
  R.~P.~Woodard,
  ``Avoiding dark energy with 1/r modifications of gravity,''
  Lect.\ Notes Phys.\  {\bf 720}, 403 (2007)
  [astro-ph/0601672].

\bibitem{Horndeski:1974wa} 
  G.~W.~Horndeski,
  ``Second-order scalar-tensor field equations in a four-dimensional space,''
  Int.\ J.\ Theor.\ Phys.\  {\bf 10}, 363 (1974).

\bibitem{Kobayashi:2011nu} 
  T.~Kobayashi, M.~Yamaguchi and J.~Yokoyama,
  ``Generalized G-inflation: Inflation with the most general second-order field equations,''
  Prog.\ Theor.\ Phys.\  {\bf 126}, 511 (2011)
  [arXiv:1105.5723 [hep-th]].

\bibitem{Tasinato:2014eka} 
  G.~Tasinato,
  ``Cosmic Acceleration from Abelian Symmetry Breaking,''
  JHEP {\bf 1404}, 067 (2014)
  [arXiv:1402.6450 [hep-th]].

\bibitem{Heisenberg:2014rta} 
  L.~Heisenberg,
  ``Generalization of the Proca Action,''
  JCAP {\bf 1405}, 015 (2014)
  [arXiv:1402.7026 [hep-th]].

\bibitem{Allys:2015sht} 
  E.~Allys, P.~Peter and Y.~Rodriguez,
  ``Generalized Proca action for an Abelian vector field,''
  JCAP {\bf 1602}, no. 02, 004 (2016)
  [arXiv:1511.03101 [hep-th]].

\bibitem{Gates:1995fx-0} 
  S.~J.~Gates, Jr.,
  ``Why auxiliary fields matter: The Strange case of the 4-D, N=1 supersymmetric QCD effective action,''
  Phys.\ Lett.\ B {\bf 365}, 132 (1996)
  [hep-th/9508153].
\bibitem{Gates:1995fx-1}
  S.~J.~Gates, Jr.,
  ``Why auxiliary fields matter: The strange case of the 4-D, N=1 supersymmetric QCD effective action. 2.,''
  Nucl.\ Phys.\ B {\bf 485}, 145 (1997)
  [hep-th/9606109].
\bibitem{Gates:1995fx-2}
  S.~J.~Gates, Jr., M.~T.~Grisaru, M.~E.~Knutt and S.~Penati,
  ``The Superspace WZNW action for 4-D, N=1 supersymmetric QCD,''
  Phys.\ Lett.\ B {\bf 503}, 349 (2001)
  [hep-ph/0012301].
\bibitem{Gates:1995fx-3}
  S.~J.~Gates, Jr., M.~T.~Grisaru, M.~E.~Knutt, S.~Penati and H.~Suzuki,
  ``Supersymmetric gauge anomaly with general homotopic paths,''
  Nucl.\ Phys.\ B {\bf 596}, 315 (2001)
  [hep-th/0009192].
\bibitem{Gates:1995fx-4}
  S.~J.~Gates, Jr., M.~T.~Grisaru and S.~Penati,
  ``Holomorphy, minimal homotopy and the 4-D, N=1 supersymmetric Bardeen-Gross-Jackiw anomaly,''
  Phys.\ Lett.\ B {\bf 481}, 397 (2000)
  [hep-th/0002045].

\bibitem{Nemeschansky:1984cd-0} 
  D.~Nemeschansky and R.~Rohm,
  ``Anomaly Constraints On Supersymmetric Effective Lagrangians,''
  Nucl.\ Phys.\ B {\bf 249}, 157 (1985).
\bibitem{Nemeschansky:1984cd-1} 
  M.~Nitta,
  ``A Note on supersymmetric WZW term in four dimensions,''
  Mod.\ Phys.\ Lett.\ A {\bf 15}, 2327 (2000)
  [hep-th/0101166].

\bibitem{Bergshoeff:1984wb} 
  E.~A.~Bergshoeff, R.~I.~Nepomechie and H.~J.~Schnitzer,
  ``Supersymmetric Skyrmions in Four-dimensions,''
  Nucl.\ Phys.\ B {\bf 249}, 93 (1985).

\bibitem{Freyhult:2003zb} 
  L.~Freyhult,
  ``The Supersymmetric extension of the Faddeev model,''
  Nucl.\ Phys.\ B {\bf 681}, 65 (2004)
  [hep-th/0310261].

\bibitem{Antoniadis:2007xc} 
  I.~Antoniadis, E.~Dudas and D.~M.~Ghilencea,
  ``Supersymmetric Models with Higher Dimensional Operators,''
  JHEP {\bf 0803}, 045 (2008)
  [arXiv:0708.0383 [hep-th]].

\bibitem{Fujimori:2016udq} 
  T.~Fujimori, M.~Nitta and Y.~Yamada,
  ``Ghostbusters in higher derivative supersymmetric theories: who is afraid of propagating auxiliary fields?,''
  JHEP {\bf 1609}, 106 (2016)
  [arXiv:1608.01843 [hep-th]].

\bibitem{Khoury:2010gb} 
  J.~Khoury, J.~L.~Lehners and B.~Ovrut,
  ``Supersymmetric P(X,$\phi$) and the Ghost Condensate,''
  Phys.\ Rev.\ D {\bf 83}, 125031 (2011)
  [arXiv:1012.3748 [hep-th]].


\bibitem{Khoury:2011da} 
  J.~Khoury, J.~L.~Lehners and B.~A.~Ovrut,
  ``Supersymmetric Galileons,''
  Phys.\ Rev.\ D {\bf 84}, 043521 (2011)
  [arXiv:1103.0003 [hep-th]].

\bibitem{Koehn:2012ar} 
  M.~Koehn, J.~L.~Lehners and B.~A.~Ovrut,
  ``Higher-Derivative Chiral Superfield Actions Coupled to N=1 Supergravity,''
  Phys.\ Rev.\ D {\bf 86}, 085019 (2012)
  [arXiv:1207.3798 [hep-th]].

\bibitem{Koehn:2012te} 
  M.~Koehn, J.~L.~Lehners and B.~Ovrut,
  ``Ghost condensate in $N=1$ supergravity,''
  Phys.\ Rev.\ D {\bf 87}, no. 6, 065022 (2013)
  [arXiv:1212.2185 [hep-th]].

\bibitem{Buchbinder:1994iw-0} 
  I.~L.~Buchbinder, S.~Kuzenko and Z.~Yarevskaya,
  ``Supersymmetric effective potential: Superfield approach,''
  Nucl.\ Phys.\ B {\bf 411}, 665 (1994).
\bibitem{Buchbinder:1994iw-1}
  I.~L.~Buchbinder, S.~M.~Kuzenko and A.~Y.~Petrov,
  ``Superfield chiral effective potential,''
  Phys.\ Lett.\ B {\bf 321}, 372 (1994).
\bibitem{Buchbinder:1994iw-2}
  A.~T.~Banin, I.~L.~Buchbinder and N.~G.~Pletnev,
  ``On quantum properties of the four-dimensional generic chiral superfield model,''
  Phys.\ Rev.\ D {\bf 74}, 045010 (2006)
  [hep-th/0606242].
\bibitem{Buchbinder:1994iw-3}
  S.~M.~Kuzenko and S.~J.~Tyler,
  ``The one-loop effective potential of the Wess-Zumino model revisited,''
  JHEP {\bf 1409}, 135 (2014)
  [arXiv:1407.5270 [hep-th]].

\bibitem{Farakos:2012qu} 
  F.~Farakos and A.~Kehagias,
  ``Emerging Potentials in Higher-Derivative Gauged Chiral Models Coupled to N=1 Supergravity,''
  JHEP {\bf 1211}, 077 (2012)
  [arXiv:1207.4767 [hep-th]].

\bibitem{Nicolis:2008in} 
  A.~Nicolis, R.~Rattazzi and E.~Trincherini,
  ``The Galileon as a local modification of gravity,''
  Phys.\ Rev.\ D {\bf 79}, 064036 (2009)
  [arXiv:0811.2197 [hep-th]].


\bibitem{Sasaki:2012ka} 
  S.~Sasaki, M.~Yamaguchi and D.~Yokoyama,
  ``Supersymmetric DBI inflation,''
  Phys.\ Lett.\ B {\bf 718}, 1 (2012)
  [arXiv:1205.1353 [hep-th]].

\bibitem{Aoki:2014pna-0} 
  S.~Aoki and Y.~Yamada,
  ``Inflation in supergravity without K\"ahler potential,''
  Phys.\ Rev.\ D {\bf 90}, no. 12, 127701 (2014)
  [arXiv:1409.4183 [hep-th]].
\bibitem{Aoki:2014pna-1} 
  S.~Aoki and Y.~Yamada,
  ``Impacts of supersymmetric higher derivative terms on inflation models in supergravity,''
  JCAP {\bf 1507}, no. 07, 020 (2015)
  [arXiv:1504.07023 [hep-th]].
  
\bibitem{Adam:2013awa-0} 
  C.~Adam, J.~M.~Queiruga, J.~Sanchez-Guillen and A.~Wereszczynski,
  ``Extended Supersymmetry and BPS solutions in baby Skyrme models,''
  JHEP {\bf 1305}, 108 (2013)
  [arXiv:1304.0774 [hep-th]].

\bibitem{Adam:2013awa-1} 
  C.~Adam, J.~M.~Queiruga, J.~Sanchez-Guillen and A.~Wereszczynski,
  ``N=1 supersymmetric extension of the baby Skyrme model,''
  Phys.\ Rev.\ D {\bf 84}, 025008 (2011)
  [arXiv:1105.1168 [hep-th]].

\bibitem{Nitta:2014pwa-0} 
  M.~Nitta and S.~Sasaki,
  ``BPS States in Supersymmetric Chiral Models with Higher Derivative Terms,''
  Phys.\ Rev.\ D {\bf 90}, no. 10, 105001 (2014)
  [arXiv:1406.7647 [hep-th]].
\bibitem{Nitta:2014pwa-1} 
  M.~Nitta and S.~Sasaki,
  ``Classifying BPS States in Supersymmetric Gauge Theories Coupled to Higher Derivative Chiral Models,''
  Phys.\ Rev.\ D {\bf 91}, 125025 (2015)
  [arXiv:1504.08123 [hep-th]].

\bibitem{Bolognesi:2014ova-0} 
  S.~Bolognesi and W.~Zakrzewski,
  ``Baby Skyrme Model, Near-BPS Approximations and Supersymmetric Extensions,''
  Phys.\ Rev.\ D {\bf 91}, no. 4, 045034 (2015)
  [arXiv:1407.3140 [hep-th]].
\bibitem{Bolognesi:2014ova-1} 
  J.~M.~Queiruga,
  ``Baby Skyrme model and fermionic zero modes,''
  Phys.\ Rev.\ D {\bf 94}, no. 6, 065022 (2016)
  [arXiv:1606.02869 [hep-th]].

\bibitem{Gudnason:2015ryh-0} 
  S.~B.~Gudnason, M.~Nitta and S.~Sasaki,
  ``A supersymmetric Skyrme model,''
  JHEP {\bf 1602}, 074 (2016)
  [arXiv:1512.07557 [hep-th]].
\bibitem{Gudnason:2015ryh-1} 
  ``Topological solitons in the supersymmetric Skyrme model,''
  JHEP {\bf 1701}, 014 (2017)
  [arXiv:1608.03526 [hep-th]].

\bibitem{Queiruga:2015xka} 
  J.~M.~Queiruga,
  ``Skyrme-like models and supersymmetry in 3+1 dimensions,''
  Phys.\ Rev.\ D {\bf 92}, no. 10, 105012 (2015)
  [arXiv:1508.06692 [hep-th]].

\bibitem{Queiruga:2017blc} 
  J.~M.~Queiruga and A.~Wereszczynski,
  ``Non-uniqueness of the supersymmetric extension of the $O(3)$ $\sigma$-model,''
  arXiv:1703.07343 [hep-th].

\bibitem{Eto:2012qda} 
  M.~Eto, T.~Fujimori, M.~Nitta, K.~Ohashi and N.~Sakai,
  ``Higher Derivative Corrections to Non-Abelian Vortex Effective Theory,''
  Prog.\ Theor.\ Phys.\  {\bf 128}, 67 (2012)
  [arXiv:1204.0773 [hep-th]].

\bibitem{Nitta:2014fca} 
  M.~Nitta and S.~Sasaki,
  ``Higher Derivative Corrections to Manifestly Supersymmetric Nonlinear Realizations,''
  Phys.\ Rev.\ D {\bf 90}, no. 10, 105002 (2014)
  [arXiv:1408.4210 [hep-th]].

\bibitem{Nitta:2017yuf-0} 
  M.~Nitta, S.~Sasaki and R.~Yokokura,
  ``Supersymmetry Breaking in Spatially Modulated Vacua,''
  arXiv:1706.05232 [hep-th].
\bibitem{Nitta:2017yuf-1} 
  M.~Nitta, S.~Sasaki and R.~Yokokura,
  ``Spatially Modulated Vacua in Relativistic Field Theories,''
  arXiv:1706.02938 [hep-th].

\bibitem{Farakos:2013zya} 
  F.~Farakos, C.~Germani and A.~Kehagias,
  ``On ghost-free supersymmetric galileons,''
  JHEP {\bf 1311}, 045 (2013)
  [arXiv:1306.2961 [hep-th]].

  \bibitem{Farakos:2012je} 
  F.~Farakos, C.~Germani, A.~Kehagias and E.~N.~Saridakis,
  ``A New Class of Four-Dimensional N=1 Supergravity with Non-minimal Derivative Couplings,''
  JHEP {\bf 1205}, 050 (2012)
    [arXiv:1202.3780 [hep-th]].
    
 \bibitem{Kimura:2016irk} 
  T.~Kimura, A.~Mazumdar, T.~Noumi and M.~Yamaguchi,
  ``Nonlocal $ \mathcal{N}=1 $ supersymmetry,''
  JHEP {\bf 1610}, 022 (2016)
  [arXiv:1608.01652 [hep-th]].
  

    
\bibitem{Cecotti:1986gb} 
  S.~Cecotti and S.~Ferrara,
  ``Supersymmetric Born-infeld Lagrangians,''
  Phys.\ Lett.\ B {\bf 187}, 335 (1987).

\bibitem{Bagger:1996wp} 
  J.~Bagger and A.~Galperin,
  ``A New Goldstone multiplet for partially broken supersymmetry,''
  Phys.\ Rev.\ D {\bf 55}, 1091 (1997)
  [hep-th/9608177].

\bibitem{Rocek:1997hi} 
  M.~Rocek and A.~A.~Tseytlin,
  ``Partial breaking of global D = 4 supersymmetry, constrained superfields, and three-brane actions,''
  Phys.\ Rev.\ D {\bf 59}, 106001 (1999)
  [hep-th/9811232].



\bibitem{Kuzenko:2002vk-0} 
  S.~M.~Kuzenko and S.~A.~McCarthy,
  ``Nonlinear selfduality and supergravity,''
  JHEP {\bf 0302}, 038 (2003)
  [hep-th/0212039].
\bibitem{Kuzenko:2002vk-1} 
  S.~M.~Kuzenko and S.~A.~McCarthy,
  ``On the component structure of N=1 supersymmetric nonlinear electrodynamics,''
  JHEP {\bf 0505}, 012 (2005)
  [hep-th/0501172].

  \bibitem{Abe:2015nxa} 
  H.~Abe, Y.~Sakamura and Y.~Yamada,
  ``Matter coupled Dirac-Born-Infeld action in four-dimensional N=1 conformal supergravity,''
  Phys.\ Rev.\ D {\bf 92}, no. 2, 025017 (2015)
    [arXiv:1504.01221 [hep-th]].

\bibitem{Cecotti:1986jy} 
  S.~Cecotti, S.~Ferrara and L.~Girardello,
  ``Structure of the Scalar Potential in General $N=1$ Higher Derivative Supergravity in Four-dimensions,''
  Phys.\ Lett.\ B {\bf 187}, 321 (1987).

\bibitem{Farakos:2013zsa} 
F.~Farakos, S.~Ferrara, A.~Kehagias and M.~Porrati,
  ``Supersymmetry Breaking by Higher Dimension Operators,''
  Nucl.\ Phys.\ B {\bf 879}, 348 (2014)
  [arXiv:1309.1476 [hep-th]].


  \bibitem{Dudas:2015vka} 
  E.~Dudas and D.~M.~Ghilencea,
    ``Effective operators in SUSY, superfield constraints and searches for a UV completion,''
  JHEP {\bf 1506}, 124 (2015)
  [arXiv:1503.08319 [hep-th]].
  
\bibitem{Kuzenko:2000tg-0} 
  S.~M.~Kuzenko and S.~Theisen,
  ``Supersymmetric duality rotations,''
  JHEP {\bf 0003}, 034 (2000)
  [hep-th/0001068].
\bibitem{Kuzenko:2000tg-1} 
  S.~M.~Kuzenko and S.~Theisen,
  ``Nonlinear selfduality and supersymmetry,''
  Fortsch.\ Phys.\  {\bf 49}, 273 (2001)
  [hep-th/0007231].

\bibitem{Kuzenko:2000tg-2} 
  S.~M.~Kuzenko,
  ``The Fayet-Iliopoulos term and nonlinear self-duality,''
  Phys.\ Rev.\ D {\bf 81}, 085036 (2010)
  [arXiv:0911.5190 [hep-th]].

\bibitem{Villasenor:2002kw} 
  E.~J.~S.~Villasenor,
  ``Higher derivative fermionic field theories,''
  J.\ Phys.\ A {\bf 35}, 6169 (2002)
  [hep-th/0203197].
  
\bibitem{Kimura:2017gcy} 
  R.~Kimura, Y.~Sakakihara and M.~Yamaguchi,
  ``Ghost free systems with coexisting bosons and fermions,''
  Phys.\ Rev.\ D {\bf 96}, no. 4, 044015 (2017)
  [arXiv:1704.02717 [hep-th]].
  
\bibitem{Wess:1992cp} 
  J.~Wess and J.~Bagger,
  ``Supersymmetry and supergravity,''
  Princeton, USA: Univ. Pr. (1992) 259 p

\bibitem{Abe:2015fha} 
  H.~Abe, Y.~Sakamura and Y.~Yamada,
  ``Massive vector multiplet inflation with Dirac-Born-Infeld type action,''
  Phys.\ Rev.\ D {\bf 91}, no. 12, 125042 (2015)
  [arXiv:1505.02235 [hep-th]].

\bibitem{Kuzenko:2011ti-0} 
  S.~M.~Kuzenko and S.~J.~Tyler,
  ``Complex linear superfield as a model for Goldstino,''
  JHEP {\bf 1104}, 057 (2011)
  [arXiv:1102.3042 [hep-th]].
\bibitem{Kuzenko:2011ti-1}
  F.~Farakos and R.~von Unge,
  ``Complex Linear Effective Theory and Supersymmetry Breaking Vacua,''
  Phys.\ Rev.\ D {\bf 91}, no. 4, 045024 (2015)
  [arXiv:1403.0935 [hep-th]].

\bibitem{Ciupke:2015msa} 
  D.~Ciupke, J.~Louis and A.~Westphal,
  ``Higher-Derivative Supergravity and Moduli Stabilization,''
  JHEP {\bf 1510}, 094 (2015)
  [arXiv:1505.03092 [hep-th]].

\bibitem{Broy:2015zba} 
  B.~J.~Broy, D.~Ciupke, F.~G.~Pedro and A.~Westphal,
  ``Starobinsky-Type Inflation from $\alpha'$-Corrections,''
  JCAP {\bf 1601}, 001 (2016)
  [arXiv:1509.00024 [hep-th]].

\bibitem{Maleknejad:2011jw-0} 
  A.~Maleknejad and M.~M.~Sheikh-Jabbari,
  ``Gauge-flation: Inflation From Non-Abelian Gauge Fields,''
  Phys.\ Lett.\ B {\bf 723}, 224 (2013)
  [arXiv:1102.1513 [hep-ph]].
\bibitem{Maleknejad:2011jw-1} 
  ``Non-Abelian Gauge Field Inflation,''
  Phys.\ Rev.\ D {\bf 84}, 043515 (2011)
  [arXiv:1102.1932 [hep-ph]].
\bibitem{Maleknejad:2011jw-2} 
  A.~Maleknejad, M.~M.~Sheikh-Jabbari and J.~Soda,
  ``Gauge Fields and Inflation,''
  Phys.\ Rept.\  {\bf 528}, 161 (2013)
  [arXiv:1212.2921 [hep-th]].
\bibitem{Bagger:1997pi} 
  J.~Bagger and A.~Galperin,
  ``The Tensor Goldstone multiplet for partially broken supersymmetry,''
  Phys.\ Lett.\ B {\bf 412}, 296 (1997)
  [hep-th/9707061].

\bibitem{Aoki:2016cnw} 
  S.~Aoki and Y.~Yamada,
  ``DBI action of real linear superfield in 4D 
$ \mathcal{N} $ = 1 conformal supergravity,''
  JHEP {\bf 1606}, 168 (2016)
  [arXiv:1603.06770 [hep-th]].
 
\end{thebibliography}
\end{document}